\documentclass{article}
\usepackage{amsmath, amssymb}
\usepackage{xcolor} 
\usepackage[authoryear]{natbib}
\usepackage{booktabs}

\renewcommand{\today}{%
  \number\day\space
  \ifcase\month\or
  January\or February\or March\or April\or May\or June\or
  July\or August\or September\or October\or November\or December
  \fi
  \space\number\year}

\title{The Cochran--Mantel--Haenszel test as a test of the average treatment effect in stratified randomized trials}

\author{%
Ekkehard Glimm\thanks{%
Novartis Pharma AG, Basel, Switzerland.%
}%
\thanks{%
Medical Faculty, Otto von Guericke University Magdeburg, Magdeburg, Germany.%
}%
\thanks{%
Corresponding author. Email: ekkehard.glimm@novartis.com%
}%
}

\date{\today}

\begin{document}

\maketitle

\begin{abstract}
This note revisits the Cochran--Mantel--Haenszel (CMH) test in light of the increasing use of causal inference methods in clinical trials. It shows that the CMH test statistic can be interpreted as a standardized version of the augmented inverse probability weighted (AIPW) estimate of the causal average treatment effect. Consequently, it is not restricted to an interpretation as a test of the common-odds ratio in a stratified logistic regression model. The note then discusses conditional and unconditional variances for this estimator as well as variance estimation and its implications for testing. The results clarify the relationship between the classical conditional CMH test and Wald-type AIPW tests, and show that, under equal treatment allocation, the CMH test is an asymptotically exact test of the weak null hypothesis of zero average treatment effect, without any restriction on the heterogeneity of stratum-specific effects. Under unequal allocation, it can be liberal or conservative. A simulation study supports these findings.
\end{abstract}


\section{Introduction}

The Cochran--Mantel--Haenszel (CMH) test \citep{MantelHaenszel1959} has been a workhorse of statistical inference for binary endpoints in clinical trials for decades. In textbooks, it is usually introduced as a test of the null hypothesis that the common odds ratio is 1 in a logistic regression model with a treatment effect and a stratum effect but no interaction, against the alternative that this common odds ratio differs from 1.

In recent years, causal inference has become more prominent in clinical trials. This has led to increased awareness of the role of covariates in the analysis of clinical trial data. Methods such as augmented inverse probability weighting (AIPW) \citep{Tsiatis2006,HernanRobins2020} are now applied in clinical trials, since it has been recognized that these methods allow unbiased estimation of treatment effects under weaker model assumptions than those required by the linear and nonlinear regression models that had been predominant before; see, for example, \citet{YeShaoYiZhao2022} and \citet{YeBannickYiShao2023}.

In this note, we revisit the CMH test from a causal inference perspective. Under stratified randomization, the numerator of the CMH statistic is an unbiased estimator of the average treatment effect (ATE). It coincides with the AIPW, IPW and standardization estimators for any outcome model using the stratification as the only factor. The question addressed here is therefore not which estimator to use, but how to standardize it for testing.

The paper is organized as follows: In Section~\ref{sec2}, we quickly review the CMH test, representing it in a form that allows the connection with AIPW-estimation. Section~\ref{sec3} establishes this connection: under stratified randomization, the AIPW estimator coincides with $\hat{\Delta}$, the estimator in the numerator of the CMH statistic, for any outcome model that depends on the stratum only, whether saturated or not and whether correctly specified or not. Section~\ref{sec4} considers different variance estimators of the AIPW point estimate and discusses their appropriateness for different situations. Section~\ref{sec5} discusses the consequences of using these different variance estimates in testing. Section~\ref{sec6} shows that under 1:1 randomization the CMH test also controls the type I error for the weak null hypothesis of zero average treatment effect. Section~\ref{sec6_sim} compares the CMH test with the Wald test based on the AIPW estimator by simulation. The paper concludes with a discussion.

\section{The CMH test as a test of weighted averages of stratum-specific risk differences}\label{sec2}

We are investigating a binary response $Y_{i}\sim Bin(1, \pi_{j_ik_i})$ and the comparison of two treatments in a clinical trial where $i$ denotes subject, \(j_i=0,1\) treatment group of subject $i$ and $k=1,\ldots, K$ enumerates the categories of a stratification factor with $k_i$ being subject $i$'s stratum. We will from now on drop the index $i$ from $j_i$, $k_i$ when possible without ambiguity. In stratum \(k\), let \(n_{jk}\) denote the number of patients assigned to treatment \(j\), and let \(r_{jk}\) denote the number of responses among those \(n_{jk}\) patients.

Furthermore, assume that stratified randomization fixes the treatment fraction within each stratum, such that
\[
\frac{n_{1k}}{n_k}=q,
\qquad
\frac{n_{0k}}{n_k}=1-q,
\qquad
n_k=n_{0k}+n_{1k},
\]
with the same allocation fraction \(q\) for all strata.

Define the observed within-stratum response proportions and the pooled within-stratum response proportion
\[
p_{1k}=\frac{r_{1k}}{n_{1k}},
\qquad
p_{0k}=\frac{r_{0k}}{n_{0k}},
\qquad
p_{\cdot k}=\frac{r_{0k}+r_{1k}}{n_k}=q\,p_{1k}+(1-q)\,p_{0k}.
\]

The Cochran--Mantel--Haenszel test statistic can then be written as
\begin{equation}\label{CMH1}
X^2_{\mathrm{CMH}}
=
\frac{
q(1-q)\left[\sum_{k=1}^K n_k(p_{1k}-p_{0k})\right]^2
}{
\sum_{k=1}^K
\frac{
n_k^2 p_{\cdot k}(1-p_{\cdot k})
}{n_k-1}
}.
\end{equation}
A slightly simpler large-sample approximation is
\[
X^2_{\mathrm{CMH}}
\approx
\frac{
q(1-q)\left[\sum_{k=1}^K n_k(p_{1k}-p_{0k})\right]^2
}{
\sum_{k=1}^K
n_k p_{\cdot k}(1-p_{\cdot k})
}.
\]

Under the strong null hypothesis $H_0: \pi_{1k}=\pi_{0k}$ for every stratum $k$, \(X^2_{\mathrm{CMH}}\) is asymptotically \(\chi^2_1\)-distributed.

\section{Connection with AIPW estimation}\label{sec3}

Let $S\in \{1,\ldots, K\}$ denote stratum and \(Z\in\{0,1\}\) treatment assignment, both interpreted as random variables. Every subject belongs to a stratum with
$
\Pr(S=k)=\psi_k, k=1,\dots,K,
$
where \(\sum_{k=1}^K \psi_k=1\).

Due to stratified randomization, we have
$
\Pr(Z=1\mid S=k)=q, \Pr(Z=0\mid S=k)=1-q,
$
the same in every stratum. Thus
$
\pi_{1k}=\Pr(Y=1\mid Z=1,S=k),
\pi_{0k}=\Pr(Y=1\mid Z=0,S=k).
$
As the ATE, we use the population-average causal risk difference which is defined as
\[
\Delta
=
\sum_{k=1}^K \psi_k(\pi_{1k}-\pi_{0k}).
\]

Considering the \(n\) subjects as a sample from a stratified population with \(n_k\) subjects from stratum \(k\), we have
\[
(n_1,\dots,n_K)\sim \mathrm{Multinomial}(n;\psi_1,\dots,\psi_K).
\]

The estimator of the marginal risk difference is
\begin{equation}\label{hatdelta}
\widehat{\Delta}
=
\sum_{k=1}^K \frac{n_k}{n}(p_{1k}-p_{0k})
=
\frac{1}{n}\sum_{k=1}^K
\left(
\frac{r_{1k}}{q}-\frac{r_{0k}}{1-q}
\right).
\end{equation}

This estimator is unbiased for \(\Delta\):
\[
E(\widehat{\Delta}\mid n_1,\dots,n_K)
=
\sum_{k=1}^K \frac{n_k}{n}(\pi_{1k}-\pi_{0k}).
\]
Taking expectation again over the multinomial distribution of \((n_1,\dots,n_K)\),
\[
E(\widehat{\Delta})
=
\sum_{k=1}^K E\left(\frac{n_k}{n}\right)(\pi_{1k}-\pi_{0k})
=
\sum_{k=1}^K \psi_k(\pi_{1k}-\pi_{0k})
=
\Delta.
\]
Comparing with equation~\eqref{CMH1}, we see that due to
$
\sum_{k=1}^K n_k(p_{1k}-p_{0k})
=
n\,\widehat{\Delta},
$
the numerator of the CMH test statistic is
\[
q(1-q)n^2\hat{\Delta}^2.
\]

Hence, under stratified randomization with common treatment fraction \(q\), the square root of the CMH test statistic can be viewed as a standardized version of the estimator of the population-average risk difference. Its numerator estimates the marginal risk difference, while its denominator estimates the corresponding null standard deviation used for hypothesis testing.

Regarding the connection with AIPW estimation, let $z_i\in \{0,1\}$ and $s_i\in \{1,\ldots,K\}$ be the realizations of $Z$ and $S$, respectively, of patient $i$.
An AIPW estimate of the treatment effect then has the form $\hat\Delta^{\mathrm{AIPW}}
=
\hat\mu_1^{\mathrm{AIPW}}-\hat\mu_0^{\mathrm{AIPW}}$ where
\[
\hat\mu_1^{\mathrm{AIPW}}
=
\frac{1}{n}\sum_{i=1}^n
\left[
\hat m_1(s_i)
+
\frac{z_i}{q}\{Y_i-\hat m_1(s_i)\}
\right]
\]
and
\[
\hat\mu_0^{\mathrm{AIPW}}
=
\frac{1}{n}\sum_{i=1}^n
\left[
\hat m_0(s_i)
+
\frac{1-z_i}{1-q}\{Y_i-\hat m_0(s_i)\}
\right].
\]
In this setup $\hat m_j(s_i)$ denotes the estimated response of a randomly selected patient in treatment group $j$ and stratum $s_i$ from some outcome model $m_j(k)$ predicting the response. The estimator is consistent under very mild regularity conditions in a causal inference framework even if the outcome model is misspecified \citep[see e.g.][]{Tsiatis2006}.

We use an outcome model with a separate response probability $m_j(k)=\pi_{jk}$ for every cell $(j,k)$. It is well known that with such a saturated model, standardization, IPW and AIPW all give the same estimate $\hat\Delta$ \citep{Rosenbaum1987,HernanRobins2020}.

Because stratified randomization assigns the same fraction $q$ of patients to treatment in every stratum, both treatment arms have exactly the same stratum composition. There is therefore no imbalance between the arms for any adjustment to correct. Irrespective of whether the outcome model is correct or not, the terms involving $\hat{m}_j$ cancel such that $\hat\Delta$ simplifies to the difference between the overall response proportions of the two arms, i.e.
\[
\hat\Delta^{\mathrm{AIPW}}
=
\hat\mu_1^{\mathrm{AIPW}}-\hat\mu_0^{\mathrm{AIPW}}
=
\frac{1}{n}\sum_{k=1}^K n_k(p_{1k}-p_{0k})
=
\hat\Delta.
\]
Stratification is only relevant to the variance used to standardize this estimate for testing, which is the subject of Sections~\ref{sec4}--\ref{sec6}.

In this note, the strata are the only covariates. The outcome model $m_j(k)$ is a function of the stratum alone. 
Additional baseline covariates could be included in $m_j$. This would lead to a non-saturated AIPW estimator for which the exact identities above no longer hold.
%, and which, if interactions with treatment are included, adds parameters, aggravating finite-sample problems. 
This is not considered here. The outcome model serves only as a device for estimating $\Delta$. We also note that the AIPW estimator has the form of the generalized regression (GREG) estimator of survey sampling \citep{SarndalSwenssonWretman1992}.

We emphasize that the exact equivalence between the CMH numerator and the AIPW estimator relies on the treatment fraction being the same in every stratum. In general, the CMH numerator is
\[
\sum_{k=1}^K\Big(r_{1k}-\frac{n_{1k}r_{\cdot k}}{n_k}\Big)
=\sum_{k=1}^K \frac{n_{1k}n_{0k}}{n_k}\,(p_{1k}-p_{0k}),
\]
i.e.\ a weighted sum of the stratum-specific risk differences with weights $n_{1k}n_{0k}/n_k=n_k q_k(1-q_k)$, $q_k=n_{1k}/n_k$. These weights reflect both the size of a stratum and how balanced the allocation within it is. Only if $q_k=q$ for all $k$ are they proportional to $n_k$, so that the numerator is proportional to $\hat\Delta$. Otherwise the CMH numerator estimates a differently weighted average of the true stratum-specific risk differences $\delta_k=\pi_{1k}-\pi_{0k}$, which coincides with $\Delta$ only if all $\delta_k$ are equal. The CMH test then remains a valid test of the strong null hypothesis, but its connection to $\Delta$ is lost. Under stratified block randomization the realized $q_k$ differ from $q$ only through incomplete final blocks, so that the discrepancy is of order $1/n$ and negligible for all practical purposes. Under simple randomization without stratification, the $q_k$ still converge to $q$, but only at rate $n^{-1/2}$. The two estimators are then both consistent for $\Delta$, but if the stratum-specific effects $\delta_k$ differ, their difference is of the same order as their standard error, so their asymptotic variances are not the same. We do not pursue this case further.

\section{Variance of $\hat{\Delta}$}\label{sec4}

This section considers $\mathrm{Var}(\hat{\Delta})$ under different degrees of conditioning. The distinction between conditional and unconditional variances of post-stratified estimators is not new. \citet{MiratrixSekhonYu2013} give a detailed treatment for general outcomes, both for post-stratification after complete randomization and for blocked designs. The expressions below are the corresponding results for a binary endpoint under stratified randomization with a common allocation fraction. We present them here because they make explicit how the denominator of the CMH statistic relates to these variances, which is the basis for Sections~\ref{sec5} and~\ref{sec6}.

\subsection*{Conditional variance given \(n_k\) and \(r_{\cdot k}=r_{0k}+r_{1k}\)}

The CMH test conditions on the observed stratum sizes \(n_k\) and the observed total number of responses in each stratum,
$
r_{\cdot k}=r_{0k}+r_{1k}.
$
Under the strong null hypothesis
\[
H_0:\pi_{1k}=\pi_{0k}\quad\text{for all }k,
\]
the conditional distribution of \(r_{1k}\) given \(n_k\) and \(r_{\cdot k}\) is hypergeometric:
\[
r_{1k}\mid n_k,r_{\cdot k}
\sim
\mathrm{Hypergeometric}(n_k,r_{\cdot k},n_{1k}).
\]
Its conditional mean and variance are
\[
E(r_{1k}\mid n_k,r_{\cdot k})
=
\frac{n_{1k}r_{\cdot k}}{n_k}
=
q\,r_{\cdot k},
\]
and
\[
\mathrm{Var}(r_{1k}\mid n_k,r_{\cdot k})
=
\frac{n_{1k}n_{0k}r_{\cdot k}(n_k-r_{\cdot k})}{n_k^2(n_k-1)}
=
\frac{q(1-q)\,r_{\cdot k}(n_k-r_{\cdot k})}{n_k-1}.
\]

Since
\[
\hat{\Delta}
=
\frac{1}{n}\sum_{k=1}^K
\left(
\frac{r_{1k}}{q}-\frac{r_{\cdot k}-r_{1k}}{1-q}
\right),
\]
and \(r_{\cdot k}\) is fixed under this conditioning, only \(r_{1k}\) is random. Rewriting
\[
\hat{\Delta}
=
\frac{1}{n}\sum_{k=1}^K
\left(
\frac{1}{q}+\frac{1}{1-q}
\right)r_{1k}
-\frac{1}{n(1-q)}\sum_{k=1}^K r_{\cdot k}
=
\frac{1}{n q(1-q)}\sum_{k=1}^K r_{1k}
-\frac{1}{n(1-q)}\sum_{k=1}^K r_{\cdot k}
\]
one obtains
\[
\mathrm{Var}(\hat{\Delta}\mid n_k,r_{\cdot k}, k=1,\dots,K)
=
\frac{1}{n^2q(1-q)}
\sum_{k=1}^K
\frac{n_k^2 p_{\cdot k}(1-p_{\cdot k})}{n_k-1}.
\]
For large \(n_k\), this is approximated by
\[
\mathrm{Var}(\hat{\Delta}\mid n_k,r_{\cdot k}, k=1,\dots,K)
\approx
\frac{1}{n^2q(1-q)}
\sum_{k=1}^K
n_k p_{\cdot k}(1-p_{\cdot k}).
\]

Hence, the CMH statistic can be written as
\[
X^2_{\mathrm{CMH}}
=
\frac{\hat{\Delta}^2}
{\mathrm{Var}(\hat{\Delta}\mid n_k,r_{\cdot k}, k=1,\dots,K)}.
\]

\subsection*{Conditional variance given \(n_k\) only}

If one conditions only on the observed stratum sizes \(n_k\), but not on the observed total responses \(r_{\cdot k}\), then under the data-generating model
\[
r_{1k}\sim \mathrm{Bin}(qn_k,\pi_{1k}),
\qquad
r_{0k}\sim \mathrm{Bin}((1-q)n_k,\pi_{0k}),
\]
independently across treatment groups within each stratum. Hence
\begin{equation}\label{condmean1}
E(\hat{\Delta}\mid n_1, \ldots, n_K)
=
\sum_{k=1}^K \frac{n_k}{n}(\pi_{1k}-\pi_{0k}),
\end{equation}
and
\begin{equation}\label{vareq1}
\mathrm{Var}(\hat{\Delta}\mid n_1, \ldots, n_K)
=
\frac{1}{n^2}
\sum_{k=1}^K
n_k
\left[
\frac{\pi_{1k}(1-\pi_{1k})}{q}
+
\frac{\pi_{0k}(1-\pi_{0k})}{1-q}
\right].
\end{equation}

Under the strong null hypothesis \(\pi_{1k}=\pi_{0k}=\pi_k\) for all \(k\), this simplifies to
\[
\mathrm{Var}(\hat{\Delta}\mid n_1, \ldots, n_K; H_0)
=
\frac{1}{n^2q(1-q)}
\sum_{k=1}^K
n_k\pi_k(1-\pi_k).
\]
A plug-in estimator replaces \(\pi_k\) by \(p_{\cdot k}\), yielding
\[
\widehat{\mathrm{Var}}(\hat{\Delta}\mid n_1, \ldots, n_K; H_0)
=
\frac{1}{n^2q(1-q)}
\sum_{k=1}^K
n_k p_{\cdot k}(1-p_{\cdot k}),
\]
which coincides with the large-sample approximation to the conditional CMH variance above.

Thus the familiar large-sample CMH test can also be seen as using the null variance of \(\hat{\Delta}\) conditional only on the stratum sizes \(n_k\).

Alternatively, one may estimate the variance conditional on \(n_k\) by plugging the observed treatment-specific proportions \(p_{1k}\) and \(p_{0k}\) directly into equation~\eqref{vareq1}. Under the null hypothesis, the null-based and general plug-in versions are asymptotically equivalent. Away from the null, however, they differ.

\subsection*{Unconditional variance}

The unconditional variance can be obtained via the law of total variance:
\[
\mathrm{Var}(\hat{\Delta})
=
E\bigl[\mathrm{Var}(\hat{\Delta}\mid n_1,\dots,n_K)\bigr]
+
\mathrm{Var}\bigl(E(\hat{\Delta}\mid n_1,\dots,n_K)\bigr).
\]

From equation~\eqref{condmean1}, 
\[
\mathrm{Var}\bigl(E(\hat{\Delta}\mid n_1,\dots,n_K)\bigr)
=
\mathrm{Var}\left(
\sum_{k=1}^K \frac{n_k}{n}\delta_k
\right),
\qquad \delta_k=\pi_{1k}-\pi_{0k}.
\]
Using $(n_1,\dots,n_K)\sim \mathrm{Multinomial}(n;\psi_1,\dots,\psi_K)$,
\[
\mathrm{Var}(n_k)=n\psi_k(1-\psi_k),
\qquad
\mathrm{Cov}(n_k,n_\ell)=-n\psi_k\psi_\ell,\quad k\neq \ell
\]
and
\[
\left(\sum_{k=1}^K \psi_k \delta_k\right)^2
=
\sum_{k=1}^K \psi_k^2 \delta_k^2
+
2\sum_{k<\ell} \psi_k \psi_\ell \delta_k \delta_\ell.
\]
It follows that
\[
\mathrm{Var}\bigl(E(\hat{\Delta}\mid n_1,\dots,n_K)\bigr)
=
\frac{1}{n}
\left[
\sum_{k=1}^K \psi_k \delta_k^2
-
\left(\sum_{k=1}^K \psi_k \delta_k\right)^2
\right].
\]

Taking expectation over the multinomial distribution (cf. equation~\eqref{vareq1}) gives
\[
E\bigl[\mathrm{Var}(\hat{\Delta}\mid n_1,\dots,n_K)\bigr]
=
\frac{1}{n}
\sum_{k=1}^K
\psi_k
\left[
\frac{\pi_{1k}(1-\pi_{1k})}{q}
+
\frac{\pi_{0k}(1-\pi_{0k})}{1-q}
\right].
\]

Combining terms,
\begin{align}
\mathrm{Var}(\hat{\Delta})
&=
\frac{1}{n}
\sum_{k=1}^K
\psi_k
\left[
\frac{\pi_{1k}(1-\pi_{1k})}{q}
+
\frac{\pi_{0k}(1-\pi_{0k})}{1-q}
\right]
+
\frac{1}{n}
\left[
\sum_{k=1}^K \psi_k \delta_k^2
-
\left(\sum_{k=1}^K \psi_k \delta_k\right)^2
\right]
\notag \\
&=
\frac{1}{n}
\sum_{k=1}^K
\psi_k
\left[
\frac{\pi_{1k}(1-\pi_{1k})}{q}
+
\frac{\pi_{0k}(1-\pi_{0k})}{1-q}
\right]
+
\frac{1}{n}\mathrm{Var}_\psi(D),
\label{varhatdelta}
\end{align}
where $\mathrm{Var}_\psi(D)$ denotes the variance of the random variable $D$ taking the value $\delta_k$ with probability $\psi_k$.

%The additional term $\frac1n\mathrm{Var}_\psi(D)$ reflects the difference between the treatment effect in the realized trial population, $\sum_k (n_k/n)d_k$, and the population-average effect $\Delta$. It corresponds to the distinction between sample and population average treatment effects in \citet{MiratrixSekhonYu2013}; see also \citet{Imbens2004}.

Compared with the variance conditional on $n_k$, the unconditional variance contains the additional term $\frac{1}{n}\mathrm{Var}_\psi(D)$, which reflects random variation in the stratum composition. It vanishes when all $\delta_k$ are equal. 

\subsection*{Choice of variance estimate for testing}

The CMH test uses the variance conditional on the observed \(n_k\) and \(r_{\cdot k}\), yielding an exact conditional test if the exact hypergeometric distribution is used, and an asymptotic $\chi^2$-test when the variance is plugged into the usual quadratic form.

By contrast, a causal inference interpretation would view the parameter \(\Delta\) as a marginal superpopulation parameter. Typically, Wald-type inference would be used to estimate the unconditional variance corresponding to repeated sampling from the superpopulation. In the present discrete-strata randomized setting, that variance is given by equation~\eqref{varhatdelta}. In practice, the unknown parameters in the formula are replaced by their sample analogues \(p_{jk}\) and \(n_k/n\).

Since the CMH numerator coincides with the AIPW estimator in the present setting, the relevant comparison is not between different point estimators but between different variance choices for the same estimator. Two distinctions should be kept separate. First, one may use a variance conditional on the observed stratum composition or an unconditional variance that also averages over random stratum composition. Second, one may estimate the variance under the null hypothesis or under the unrestricted model. The classical CMH test uses a conditional null-based variance, whereas Wald-type inference for AIPW estimators usually uses an unrestricted variance, often in its unconditional superpopulation form.

In the binary case, the null-based pooled variance is generally larger than the unrestricted plug-in variance away from the null. For example, for balanced randomization \(q=1/2\), the large-sample null-based variance conditional on \(n_k\) exceeds the unrestricted conditional variance by
\[
\frac{1}{n^2}\sum_{k=1}^K n_k(p_{1k}-p_{0k})^2.
\]
Thus, null-based variance estimation may reduce the value of the test statistic under alternatives. This suggests a potential power advantage for tests based on the unrestricted plug-in variance estimator. In practice, however, that advantage is often tempered by the estimator's poor finite-sample behavior (see also Section~\ref{sec6_sim}). In small or moderately sized samples, with many strata, or when some stratum-by-treatment cells are sparse or have response probabilities close to 0 or 1, the factors \(p_{jk}(1-p_{jk})\) become too small and hence tend to underestimate the corresponding variance components. Moreover, the estimated between-stratum component in the unconditional variance is unreliable when it is based on few strata. The resulting Wald tests will therefore be anti-conservative in many cases. Null-based or pooled variance estimators are often preferred for testing in comparative binomial problems precisely because they are more stable in such settings, even though they are larger than the unrestricted plug-in variance away from the null; see, for example, \citet{FarringtonManning1990}, \citet{FleissLevinPaik2003}, and \citet{Agresti2013}.

\section{Consequences for testing}\label{sec5}

Section~\ref{sec4} clarifies the role of conditioning in tests based on \(\hat{\Delta}\). If inference is carried out conditionally on the observed stratum sizes, variation due to random stratum composition is removed. In the unrestricted variance decomposition, the difference between the unconditional variance and the variance conditional on \(n_k\) is $\frac{1}{n}\mathrm{Var}_{\psi}(D)$.
This term vanishes only if the stratum-specific risk differences are all equal. Thus, conditional inference removes a source of variation that is unrelated to the treatment comparison within the realized stratum composition of the trial. For hypothesis testing, this conditional perspective has a natural appeal. If a test has a conditional type I error of at most \(\alpha\) for every value of the conditioning statistic, then it also has unconditional type I error of at most \(\alpha\), since
\[
P_{H_0}(\text{reject})
=
E_{H_0}\{P_{H_0}(\text{reject}\mid \text{conditioning statistic})\}
\leq \alpha.
\]
Thus, conditioning on the observed stratum composition does not compromise type I error control under the strong null hypothesis $\delta_k=0$ for all $k$. Rather, it eliminates from the test statistic the extra variation due to random stratum sizes.
%it avoids incorporating into the test statistic variation due solely to which strata happened to be represented in the realized sample. 
This is particularly attractive in randomized clinical trials, where the stratum composition is observed before the outcome analysis and the primary comparison is between randomized treatment groups within the realized trial population.

This argument is distinct from the question whether the variance is estimated under the null or under the unrestricted model. The classical CMH statistic combines both features: it conditions on the observed \(n_k\) and \(r_{\cdot k}\), and it uses a null-based pooled variance. An unrestricted AIPW Wald test, in contrast, typically uses an unrestricted variance estimator and may also use an unconditional superpopulation variance. A direct comparison between the two therefore mixes two effects: conditioning tends to remove variation from random stratum composition, whereas null-based pooling can increase the denominator under alternatives. At $q=1/2$, these two effects almost cancel: the null-based pooled variance exceeds the unrestricted conditional variance by $\frac1n\sum_k\psi_k\delta_k^2=\frac1n\{\mathrm{Var}_\psi(D)+\Delta^2\}$, whereas the unconditional variance exceeds it by 
$\frac1n\mathrm{Var}_\psi(D)$. The CMH denominator is therefore larger by only $\Delta^2/n$, which is of second order under local alternatives. The simulations in Section~\ref{sec6_sim} are in line with these theoretical expectations.

Consequently, the main point is not that the CMH test must have uniformly higher power than an unrestricted AIPW Wald test. Under strict Pitman local alternatives the differences discussed above are of second order, and under fixed alternatives both tests are consistent. Rather, the point is that the CMH test is a valid test based on the same stratified ATE estimator, without requiring the common-odds-ratio model often used to motivate it. It controls the type I error under the strong null hypothesis for any allocation ratio and, as shown in Section~\ref{sec6}, under the weak null hypothesis $\Delta=0$ when allocation is 1:1.

\section{Asymptotic type I error for different versions of null hypotheses}\label{sec6}

First, consider the one-sided CMH test (comparing the root of the test statistic, signed by $\hat\Delta$, with the corresponding $N(0,1)$-quantile) as an asymptotic level-$\alpha$ test of the strong null hypothesis $\delta_k=0$ for all $k$, conditionally on the observed stratum sizes and hence, by the argument of Section~\ref{sec5}, also unconditionally. It can be re-interpreted as a test of the composite null hypothesis ``no benefit in any stratum'',
\[
H_0: \delta_k\le 0 \text{ for all }k,
\]
against the restricted alternative ``no harm in any stratum and benefit in at least one stratum'',
\[
H_A: \delta_k\ge 0 \text{ for all }k,\ \text{with strict inequality for at least one }k.
\]
Within this $H_0$, $\delta_k=0$ for all $k$ is the least favourable configuration: there the rejection probability converges to $\alpha$, while at every other point some $\delta_k<0$, hence $\Delta<0$, and the rejection probability converges to $0$. The one-sided CMH test is therefore an asymptotic level-$\alpha$ test of this pair of hypotheses for any allocation ratio $q$.

This interpretation excludes configurations with benefit in some strata and harm in others, in which effects of opposite sign can cancel such that $\Delta=0$. This motivates the weak null hypothesis
\[
H_0: \Delta=\sum_{k=1}^K \psi_k \delta_k=0.
\]
Under this weak null, the conditioning argument of Section~\ref{sec5} is no longer available: given the realized stratum sizes, 
$\sum_k n_k \delta_k=\sum_k (n_k-n\psi_k)\delta_k$ is in general not zero. We therefore derive the unconditional asymptotic distribution of the CMH statistic, treating $(n_1,\dots,n_K)$ as multinomial.

Let
\[
T_n := \sum_{k=1}^K n_k(p_{1k}-p_{0k}) = n\hat\Delta,
\qquad
v_{jk}:=\pi_{jk}(1-\pi_{jk}).
\]
By equation~\eqref{varhatdelta}, under the weak null, where $\mathrm{Var}_\psi(D)=\sum_k\psi_k \delta_k^2$,
\[
\frac{T_n}{\sqrt{n}}
\overset{d}{\longrightarrow}
N\bigl(0,\ \sigma^2\bigr),
\qquad
\sigma^2
=
\sum_{k=1}^K \psi_k
\left[
\frac{v_{1k}}{q}
+
\frac{v_{0k}}{1-q}
\right]
+
\sum_{k=1}^K \psi_k \delta_k^2 .
\]
For the denominator of the CMH statistic, $p_{\cdot k}\to \bar\pi_k:=q\pi_{1k}+(1-q)\pi_{0k}$ in probability, and
\[
\frac{1}{n}\sum_{k=1}^K n_k p_{\cdot k}(1-p_{\cdot k})
\overset{p}{\longrightarrow}
\sum_{k=1}^K \psi_k \bar\pi_k(1-\bar\pi_k)
=
\sum_{k=1}^K \psi_k
\bigl[
q\,v_{1k}
+
(1-q)\,v_{0k}
\bigr]
+
q(1-q)\sum_{k=1}^K \psi_k \delta_k^2 ,
\]
where we used the identity
$\bar\pi_k(1-\bar\pi_k)=q\,v_{1k}+(1-q)\,v_{0k}+q(1-q)\delta_k^2$.
The same limit applies to the finite-sample version in equation~\eqref{CMH1} with the factors $n_k/(n_k-1)$. Hence, using $n_{1k}=qn_k$,
\[
X^2_{\mathrm{CMH}}
=
\frac{q(1-q)\,T_n^2/n}
{\left[\sum_{k=1}^K n_k p_{\cdot k}(1-p_{\cdot k})\right]/n}
\overset{d}{\longrightarrow}
c\cdot \chi_1^2,
\]
with
\begin{equation}\label{cfactor}
c
=
\frac{
\sum_{k=1}^K \psi_k\bigl[(1-q)\,v_{1k} + q\,v_{0k}\bigr]
+
q(1-q)\sum_{k=1}^K \psi_k \delta_k^2
}{
\sum_{k=1}^K \psi_k\bigl[q\,v_{1k} + (1-q)\,v_{0k}\bigr]
+
q(1-q)\sum_{k=1}^K \psi_k \delta_k^2
}.
\end{equation}
Numerator and denominator differ only by
\[
(1-2q)\sum_{k=1}^K \psi_k\,(v_{1k}-v_{0k}).
\]
Two conclusions follow immediately:
\begin{enumerate}
\item Under balanced allocation $q=\tfrac12$, $c=1$ for every configuration of $\pi_{jk}$ satisfying the weak null. The CMH test is therefore an asymptotically exact level-$\alpha$ test of the weak null hypothesis $\Delta=0$. The null-based pooled variance in its denominator exceeds the within-stratum variance by exactly the amount of between-stratum variation $\frac14\sum_k\psi_k \delta_k^2$ that the random stratum composition adds to the variance of $\hat\Delta$.
\item Under unequal allocation $q\neq\tfrac12$, $c\neq 1$ in general. The test is liberal ($c>1$) if $(1-2q)\sum_k\psi_k(v_{1k}-v_{0k})>0$, i.e.\ if the larger treatment arm has, on average, the smaller binomial variance, and conservative otherwise.
\end{enumerate}
This parallels the classical result that pooled-variance two-sample tests are robust to variance heterogeneity only under equal allocation \citep{Scheffe1970}; see also \citet{BugniCanayShaikh2018} for an analogous result under covariate-adaptive randomization.

If the nominal critical value is $\chi^2_{1,1-\alpha}$, the asymptotic rejection probability is
\[
P(X^2_{\mathrm{CMH}}>\chi^2_{1,1-\alpha})
\to
P(c\chi_1^2>\chi^2_{1,1-\alpha})
=
2\left[
1-\Phi\left(\sqrt{\chi^2_{1,1-\alpha}/c}\right)
\right].
\]
To illustrate the magnitude of deviations from nominal $\alpha$ in extreme cases, consider $K=4$ strata with $\psi_k=0.25$, $\pi_{0k}=0.3$ for all $k$, $\pi_{11}=\pi_{13}=0.55$ and $\pi_{12}=\pi_{14}=0.05$, such that $\Delta=0$ with $\delta_k=\pm 0.25$. At $\alpha=0.05$, the asymptotic rejection probability is $5\%$ for $q=0.5$, about $2.5\%$ for $q=0.1$ and about $8.7\%$ for $q=0.9$. For practically more relevant, less extreme settings, the finite-sample accuracy of the asymptotic rejection probabilities is examined in Section~\ref{sec6_sim}.

\section{Simulation}\label{sec6_sim}

\subsection*{Design}

We compare the CMH test with the Wald test based on the same estimator $\hat\Delta$ in trials with $n=100$ or $n=200$ patients, $K=4$ or $K=8$ strata, and allocation 1:1 ($q=1/2$) or 2:1 ($q=2/3$). The stratum probabilities and control response rates are
\begin{align*}
K=4:&\quad \psi=(0.40,0.30,0.20,0.10),\\
    &\quad \pi_{0\cdot}=(0.30,0.40,0.20,0.50),\\[4pt]
K=8:&\quad \psi=(0.25,0.20,0.15,0.12,0.10,0.08,0.06,0.04),\\
    &\quad \pi_{0\cdot}=(0.30,0.40,0.20,0.50,0.35,0.25,0.45,0.15).
\end{align*}
The unequal $\psi_k$ were chosen such that some strata are small, in particular for $K=8$ and $n=100$, where the smallest stratum has an expected size of 4 patients.

The response probabilities in the treatment arm are $\pi_{1k}=\pi_{0k}+\delta_k$ where $\delta_k$ is determined as follows:
\begin{itemize}
\item Strong null: $\delta_k=0$ for all $k$.
\item Weak null: $\delta_k=0.12\,s_k$ with  $s_k=(-1)^{k+1}-\sum_\ell\psi_\ell(-1)^{\ell+1}$. Hence, $s_k$ alternates in sign and is shifted such that $\sum_k\psi_k s_k=0$. For $K=4$, $\psi=(0.40,0.30,0.20,0.10)$ this gives $\delta_k=0.096,-0.144,0.096,-0.144$, respectively.
\item Power, homogeneous: $\delta_k=0.18$ for all $k$.
\item Power, heterogeneous: $\delta_k=0.18+0.12\,s_k$ with the same $s_k$ as for the weak null.
\end{itemize}
The effect $\Delta=0.18$ gives a power of about 75\% for $n=200$ and 1:1 allocation.

In each simulated trial, the stratum sizes $n_1,\dots,n_K$ were drawn from the multinomial distribution, the treatment arm in stratum $k$ received $n_{1k}=\mathrm{round}(q\,n_k)$ patients, mimicking stratified block randomization, and the numbers of responders were drawn from the binomial distributions. No restrictions are imposed, so that empty strata and strata in which one arm is empty (in particular strata with a single patient) occur. There were 100{,}000 replications per setting. The Monte Carlo standard error of a rejection rate is about 0.05 percentage points at 2.5\% and about 0.16 percentage points at 50\%.

The CMH test uses statistic~\eqref{CMH1} without continuity correction. The Wald test uses $\hat\Delta/\hat V^{1/2}$ with the plug-in version of the unconditional variance~\eqref{varhatdelta},
\[
\hat V
=
\frac{1}{n}
\sum_{k=1}^K
\frac{n_k}{n}
\left[
\frac{p_{1k}(1-p_{1k})}{q}
+
\frac{p_{0k}(1-p_{0k})}{1-q}
\right]
+
\frac{1}{n}
\left[
\sum_{k=1}^K \frac{n_k}{n}(p_{1k}-p_{0k})^2
-
\hat\Delta^2
\right].
\]
Both tests are carried out one-sided at the nominal level of 2.5\%, as usual for superiority trials: for the CMH test, the signed root $\mathrm{sign}(\hat\Delta)\,(X^2_{\mathrm{CMH}})^{1/2}$, and for the Wald test $\hat\Delta/\hat V^{1/2}$, is compared with $z_{0.975}=1.96$.

To separate the power of the two tests from their size, we also report size-adjusted power: for each combination of $K$, $n$ and $q$, the empirical 97.5\% quantile of each signed statistic in the strong-null scenario was used as critical value in the two power scenarios. This is a benchmark for a fair power comparison only. It is not a practical remedy for the liberality of the Wald test, since the critical value depends on the unknown $\psi_k$ and $\pi_{jk}$ and, under the weak null, on the unknown configuration of the $\delta_k$. To check that the results do not depend on control proportions being mostly below $0.5$, we repeated all settings with mirrored proportions, $\pi_{0k}\to 1-\pi_{1k}$ and $\pi_{1k}\to 1-\pi_{0k}$, which leaves all $\delta_k$ unchanged.

\paragraph{Empty strata and empty arms.}
A stratum in which one arm is empty contributes nothing to the CMH statistic: its weight $n_{1k}n_{0k}/n_k$ in the numerator and its contribution to the variance are both zero. In our implementation such strata are simply omitted, which leaves the statistic unchanged. For the Wald test, in contrast, no estimate $p_{1k}-p_{0k}$ is available for such a stratum. We dropped these strata and rescaled the weights $n_k/n$ of the remaining strata to sum to 1. Strictly speaking, this changes the estimand to the average treatment effect over the strata in which both arms are observed, which is somewhat awkward, since a stratum is then excluded merely because it happened to be represented by too few patients. Empty strata ($n_k=0$) receive weight 0 in both tests and are irrelevant. With $n=100$ and $K=8$, about 10\% of the simulated trials contain at least one empty stratum or a stratum with an empty arm (Table~\ref{tab:sim_CMH_Wald}, column Drop.). Otherwise, this is rare.

\subsection*{Results}

The results are shown in Table~\ref{tab:sim_CMH_Wald}. The table contains the upper-tail rates for the original settings only. Lower-tail rates and the results for the mirrored settings reported in the text are given in the Supplementary Material.

\emph{Type I error.} The CMH test keeps the one-sided nominal level of 2.5\% well in all settings, with rejection rates between 2.30\% and 2.62\%, under both the strong and the weak null hypothesis, even for $K=8$ and $n=100$ with strata of only a few patients. This confirms the well-known observation that the asymptotic CMH test does not require large strata, but only a sufficiently large total sample size; see, for example, \citet{MantelFleiss1980}. The Wald test is liberal in all settings, with rejection rates between 2.69\% and 4.03\%. The excess is largest with more strata, smaller samples and unequal allocation, i.e.\ when stratum-by-treatment cells are small, and it decreases with increasing $n$. This is in line with the arguments at the end of Section~\ref{sec4}: the factors $p_{jk}(1-p_{jk})$ in small cells tend to underestimate the corresponding variance components.

\emph{Asymmetry under unequal allocation.} Under 1:1 allocation, the rejection rates of both tests in the upper and the lower tail are practically equal. Under 2:1 allocation, the Wald test is clearly more liberal in the upper tail, which is the relevant one for superiority testing (3.10--4.03\%), than in the lower tail (2.49--3.25\%). The reason is that the smaller control arm dominates the variance, with weight $1/(1-q)=3$ against $1/q=1.5$ for the treatment arm. Since most $\pi_{0k}$ are below $0.5$, a control proportion $p_{0k}$ that happens to be low increases $\hat\Delta$ and at the same time decreases $p_{0k}(1-p_{0k})$ and hence $\hat V$. Large positive values of the Wald statistic are therefore inflated, while negative values are not. Under 1:1 allocation, the two arms contribute symmetrically and this effect largely cancels. The CMH test, whose pooled variance does not depend on the arm-specific proportions, shows only a slight asymmetry in the opposite, conservative direction (upper tail 2.30--2.46\%, lower tail 2.48--2.73\% under 2:1 allocation). In the mirrored settings, both asymmetries are reversed: for the Wald test, upper tail 2.54--3.28\% and lower tail 3.07--4.01\%; for the CMH test, upper tail 2.49--2.67\% and lower tail 2.26--2.44\%.

\emph{Weak null hypothesis.} The simulated rejection rates of the CMH test under the weak null agree with the asymptotic rates obtained from equation~\eqref{cfactor} for the same settings (column As.). For 1:1 allocation, the asymptotic rate is exactly 2.5\%. The simulated rates are 2.40--2.48\%. For 2:1 allocation, the asymptotic rates are 2.46\% ($K=4$) and 2.55\% ($K=8$), and the simulated rates are 2.30--2.46\%, reflecting in addition the slight conservativeness of the upper tail noted above. With the moderate heterogeneity used here, the deviation of $c$ from 1 is small. The example at the end of Section~\ref{sec6} shows that it can be considerably larger for stronger heterogeneity and more extreme allocation.

\emph{Power.} With nominal critical values, the Wald test has higher power than the CMH test, by up to 7.7 percentage points ($K=8$, $n=100$, 2:1). This difference is due to the inflated type I error of the Wald test. After size adjustment, the power of the two tests is practically identical under 1:1 allocation (differences of at most 0.3 percentage points), and the CMH test is ahead by 1.5 to 2.7 percentage points (if the CMH test critical value is adjusted downwards) or 0.6 to 1.6 percentage points (if the nominal level is used for CMH) under 2:1 allocation. The result for 1:1 allocation agrees with Section~\ref{sec5}: the null-based pooled variance of the CMH test exceeds the unconditional variance used by the Wald test by only $\Delta^2/n$, which is negligible here. Under unequal allocation, the pooled variance in addition weights the arm-specific binomial variances $v_{1k}$ and $v_{0k}$ with $q$ and $1-q$ instead of $1-q$ and $q$, which asymptotically gives the Wald test a slight advantage in our power settings, where the larger treatment arm mostly has the larger binomial variance. In finite samples, this is outweighed by the upward skewness of the Wald statistic, which requires a considerably larger size-adjusted critical value (2.07--2.21 instead of 1.96). The potential power advantage of estimating the variance without restriction, discussed in Section~\ref{sec4}, is therefore small in these settings, and its sign depends on the configuration: in the mirrored settings, the size-adjusted power under 2:1 allocation is slightly higher for the Wald test (by 0.5 to 1.0 percentage points). Hence, neither test has a consistent power advantage. The robust difference between the two tests is the type I error.

\emph{Heterogeneity.} Heterogeneous treatment effects reduce the power of both tests by at most about one percentage point compared with homogeneous effects of the same $\Delta$. This reflects the additional variance term $\frac1n\mathrm{Var}_\psi(D)$, which is small compared with the within-stratum variance for the heterogeneity considered here.

\emph{Two-sided testing.} For two-sided tests at the 5\% level (not shown in detail), the CMH test again keeps the level (4.79--5.19\%), and the Wald test is liberal (5.36--7.24\%). Since a two-sided test counts both tails, the asymmetry of the Wald test is partly hidden: its two-sided rate overstates the lower-tail error and understates the upper-tail error. Accordingly, two-sided size adjustment does not fully correct the upper tail, and the Wald test then even appears slightly more powerful under 2:1 allocation (by 0.2--1.1 percentage points). For one-sided superiority testing, this comparison is misleading.

In summary, the CMH test controls the type I error in all settings studied, without relevant loss of power, in line with Sections~\ref{sec5} and~\ref{sec6}.

\begin{table}[!htbp]
\centering
\caption{Simulated one-sided rejection rates (\%, upper tail, nominal level 2.5\%) of the CMH test and the Wald test for the stratified ATE estimator; 100{,}000 replications per setting.}
\label{tab:sim_CMH_Wald}
\footnotesize
\setlength{\tabcolsep}{4pt}
\begin{tabular}{rrllrrrrrrrr}
\toprule
 & & & & & \multicolumn{2}{c}{Nominal} & \multicolumn{2}{c}{Size-adj.} & \multicolumn{2}{c}{Crit.\ values} & \\
\cmidrule(lr){6-7}\cmidrule(lr){8-9}\cmidrule(lr){10-11}
$K$ & $n$ & Alloc. & Scenario & As. & CMH & Wald & CMH & Wald & CMH & Wald & Drop. \\
\midrule
4 & 100 & 1:1 & strong null & 2.50 &  2.42 &  2.94 &  ---  &  ---  & 1.95 & 2.03 &  0.03 \\
  &     &     & weak null   & 2.50 &  2.46 &  2.93 &  ---  &  ---  &      &      &  0.04 \\
  &     &     & power, hom. & ---  & 46.5  & 48.7  & 47.0  & 46.9  &      &      &  0.04 \\
  &     &     & power, het. & ---  & 45.8  & 47.7  & 46.4  & 46.2  &      &      &  0.04 \\[2pt]
4 & 100 & 2:1 & strong null & 2.50 &  2.36 &  3.73 &  ---  &  ---  & 1.94 & 2.16 &  0.03 \\
  &     &     & weak null   & 2.46 &  2.30 &  3.63 &  ---  &  ---  &      &      &  0.03 \\
  &     &     & power, hom. & ---  & 41.1  & 46.5  & 42.1  & 39.5  &      &      &  0.03 \\
  &     &     & power, het. & ---  & 40.0  & 45.8  & 41.2  & 38.7  &      &      &  0.04 \\[2pt]
4 & 200 & 1:1 & strong null & 2.50 &  2.62 &  2.87 &  ---  &  ---  & 1.98 & 2.02 &  0.00 \\
  &     &     & weak null   & 2.50 &  2.40 &  2.69 &  ---  &  ---  &      &      &  0.00 \\
  &     &     & power, hom. & ---  & 75.3  & 76.1  & 74.8  & 74.8  &      &      &  0.00 \\
  &     &     & power, het. & ---  & 74.7  & 75.2  & 73.8  & 73.8  &      &      &  0.00 \\[2pt]
4 & 200 & 2:1 & strong null & 2.50 &  2.39 &  3.12 &  ---  &  ---  & 1.94 & 2.07 &  0.00 \\
  &     &     & weak null   & 2.46 &  2.31 &  3.10 &  ---  &  ---  &      &      &  0.00 \\
  &     &     & power, hom. & ---  & 69.6  & 72.1  & 70.2  & 68.5  &      &      &  0.00 \\
  &     &     & power, het. & ---  & 68.4  & 71.4  & 69.3  & 67.8  &      &      &  0.00 \\
\midrule
8 & 100 & 1:1 & strong null & 2.50 &  2.48 &  3.26 &  ---  &  ---  & 1.95 & 2.08 & 10.4  \\
  &     &     & weak null   & 2.50 &  2.48 &  3.19 &  ---  &  ---  &      &      & 10.4  \\
  &     &     & power, hom. & ---  & 45.9  & 50.6  & 46.2  & 46.1  &      &      & 10.3  \\
  &     &     & power, het. & ---  & 45.7  & 50.3  & 46.0  & 45.7  &      &      & 10.5  \\[2pt]
8 & 100 & 2:1 & strong null & 2.50 &  2.31 &  4.00 &  ---  &  ---  & 1.93 & 2.21 & 10.5  \\
  &     &     & weak null   & 2.55 &  2.41 &  4.03 &  ---  &  ---  &      &      & 10.5  \\
  &     &     & power, hom. & ---  & 41.2  & 48.9  & 42.5  & 40.2  &      &      & 10.4  \\
  &     &     & power, het. & ---  & 41.3  & 48.4  & 42.5  & 39.8  &      &      & 10.4  \\[2pt]
8 & 200 & 1:1 & strong null & 2.50 &  2.58 &  2.94 &  ---  &  ---  & 1.97 & 2.03 &  0.28 \\
  &     &     & weak null   & 2.50 &  2.45 &  2.79 &  ---  &  ---  &      &      &  0.25 \\
  &     &     & power, hom. & ---  & 75.5  & 77.2  & 75.1  & 75.1  &      &      &  0.26 \\
  &     &     & power, het. & ---  & 75.3  & 77.0  & 74.9  & 74.8  &      &      &  0.27 \\[2pt]
8 & 200 & 2:1 & strong null & 2.50 &  2.40 &  3.33 &  ---  &  ---  & 1.94 & 2.10 &  0.30 \\
  &     &     & weak null   & 2.55 &  2.46 &  3.29 &  ---  &  ---  &      &      &  0.26 \\
  &     &     & power, hom. & ---  & 70.2  & 73.6  & 70.7  & 69.2  &      &      &  0.31 \\
  &     &     & power, het. & ---  & 69.8  & 73.0  & 70.4  & 68.6  &      &      &  0.27 \\
\bottomrule
\end{tabular}

\smallskip
\begin{minipage}{0.95\textwidth}
\footnotesize
Alloc.: allocation ratio treatment:control. As.: asymptotic one-sided rejection rate of the CMH test from Section~\ref{sec6}. Nominal: rejection rates with the critical value $z_{0.975}=1.96$. Size-adj.: power with the critical values in the columns Crit.\ values, i.e.\ the empirical 97.5\% quantiles of the signed statistics under the strong null for the same $K$, $n$ and allocation. Drop.: percentage of trials with at least one empty stratum or one stratum with an empty arm.
\end{minipage}
\end{table}

\clearpage
\section{Discussion}\label{sec7}

In this note, we showed that the Cochran--Mantel--Haenszel test for the stratified comparison of a binary response to two treatments can be viewed as a test of the causal ATE in a study with covariate adjustment for one or several categorical covariates for which stratified randomization is feasible. If more than one such covariate exists, they can be combined into a single one by taking the product of their categories. We further showed that the CMH test statistic can be interpreted as a standardized version of an estimate of the ATE that is unbiased by randomization alone and coincides with the AIPW estimate, irrespective of the outcome model used.

Regarding variance estimation and the closely related power of corresponding tests, we showed that the CMH test
\begin{enumerate}
\item preserves the type I error under a strong null hypothesis of no treatment effect in any stratum, either asymptotically or exactly when the exact conditional version is used,
\item is an asymptotically exact test of the weak null hypothesis of zero average treatment effect in the practically important case of equal randomization to the two treatments, irrespective of how the treatment effect varies across strata. Under unequal randomization, it can be liberal or conservative, although in the realistic settings of Section~\ref{sec6_sim} the deviation from the nominal level is small.
\end{enumerate}

Of these observations, the first has long been well known. Nevertheless, it is useful to recall that the CMH test is valid under substantially weaker assumptions than suggested by the common textbook presentation as a test of a common odds ratio in a logistic regression model without treatment-by-stratum interaction.

%The second statement about the weak null hypothesis arises because, under equal allocation, the null-based pooled variance in the CMH denominator exceeds the average of the within-stratum binomial variances by exactly the between-stratum variation of the treatment effect, so that no separate estimate of that component is needed.

The second statement about the weak null hypothesis arises because, under equal allocation, the null-based pooled variance in the CMH denominator automatically includes the between-stratum variation of the treatment effect, so that no separate estimate of that component is needed.

The comparison with AIPW also highlights an important practical distinction between point estimation and testing. Modern covariate-adjusted estimators, including AIPW-type estimators proposed for randomized trials target marginal treatment effects and can improve the precision of point estimation relative to unadjusted estimators. However, the large-sample variance theory underlying these methods is often fragile in finite samples. When the number of strata is large relative to the sample size, or when there is substantial heterogeneity across strata, some stratum-by-treatment cells will be small or have observed response proportions close to 0 or 1. The unrestricted plug-in variance then becomes unstable and will usually underestimate the true variability, leading to liberal Wald-type tests.,

This issue is distinct from the causal interpretation of the point estimator. Under stratified randomization, the CMH statistic and the AIPW Wald statistic share the same numerator $\hat\Delta$, and under equal allocation and the weak null hypothesis their variance estimates also converge to the same limit, the true variance of $\hat\Delta$. The remaining difference is that between a score-type and a Wald-type test of the same parameter: the CMH test estimates the variance under the null, pooling treatment arms within strata, whereas the AIPW Wald test estimates it without restriction, cell by cell, and must in addition estimate the between-stratum variation of the treatment effect from the $K$ observed strata. The two tests therefore have the same first-order asymptotic properties but can behave noticeably differently in finite samples with sparse cells or response proportions near 0 or 1. For hypothesis testing, it is therefore the choice of variance estimator, not of the point estimator, that matters, and asymptotic equivalence does not guarantee comparable type I error control in finite samples. The simulations in Section~\ref{sec6_sim} illustrate this. The CMH test keeps the nominal level in all settings and in both tails, whereas the Wald test is liberal throughout. Under unequal allocation, the excess type I error of the Wald test is concentrated in one tail, depending on whether the control proportions are mostly below or above $0.5$. A two-sided analysis partly hides this. The power advantage of the Wald test at nominal critical values is due to its inflated type I error. After size adjustment, neither test has a consistent power advantage. Hence, in the settings studied, the CMH test gives up no relevant power in exchange for reliable type I error control.

%AIPW estimators can also adjust for additional baseline covariates, including continuous ones. However, asymptotic efficiency gains are guaranteed only if the working model includes the interactions of all covariates with treatment, and the same applies to any interactions among covariates, which then require three-way interactions with treatment. The stratified estimator $\hat{\Delta}$  is the limiting case of this approach: a saturated model for the stratification factors, including all interactions with treatment. Each additional covariate or interaction therefore moves the analysis towards the sparse-data situation in which, as shown above, unrestricted variance estimation becomes unreliable. In practice, most of the attainable gain usually comes from a few strongly prognostic covariates.

The finite-sample behavior of the Wald test can be improved in various ways, for example by small-sample corrections of the variance estimator and the degrees of freedom \citep{BellMcCaffrey2002}, by adding pseudo-observations to sparse cells \citep{AgrestiCaffo2000}, by bootstrapping, or by calibrating the critical value under the null hypothesis. A systematic comparison of such variants is beyond the scope of this note. We note, however, that these corrections add choices that must be prespecified and in general control the type I error only approximately. With sparse cells, bootstrap resamples suffer from the same empty arms and boundary proportions as the original data. Calibration under the strong null hypothesis by permuting treatment labels within strata, keeping the stratum sizes and the response totals fixed, leads to the exact conditional distribution of $\hat\Delta$, i.e.\ to the exact CMH test. In this sense, the CMH test is itself the natural stabilized version of a test based on $\hat\Delta$, and it is available in closed form in all standard software.

%If one wishes to interpret the CMH procedure as a test of the weak null hypothesis of zero average treatment effect, this is most natural against a restricted alternative such as ``no harm in any stratum and benefit in at least one stratum''. We agree that such restricted alternatives may appear awkward and can have undesirable properties; see, for example, \citep{SilvapulleSen2005}.
%However, avoiding this restriction does not yield an obviously more meaningful target. In clinical trials, strata are very rarely a random sample from a well-defined super-population; their composition reflects design choices, geography, eligibility criteria, and patterns of participation. Therefore, when stratum-specific effects differ in sign, the weak null $\Delta=0$ does not represent the absence of a clinically relevant treatment effect. A Wald test of $\Delta$ inherits this interpretational difficulty and, in finite stratified samples, may also have unstable operating characteristics. Thus, the restricted alternative used to motivate the CMH procedure is a more transparent approximation to the scientific question of interest.

Which null hypothesis does the CMH test actually test? Under equal allocation, it is a valid test of the weak null hypothesis $\Delta=0$, without any restriction on how the treatment effect varies across strata. For any allocation ratio, the one-sided test is also a valid test of ``no benefit in any stratum'' against ``no harm in any stratum and benefit in at least one stratum''. Such restricted alternatives may appear awkward and can have undesirable properties \citep{SilvapulleSen2005}. However, the weak null of zero average treatment effect is not necessarily the clinically more relevant hypothesis. In clinical trials, strata are very rarely a random sample from a well-defined super-population. Their composition reflects design choices, geography, eligibility criteria, and patterns of participation. If stratum-specific effects differ in sign, $\Delta=0$ merely states that benefit and harm cancel for this particular stratum composition, not that the treatment is without effect. In such a situation, neither a test of the weak null nor a test against the restricted alternative answers a relevant clinical question. The heterogeneity itself would have to be addressed.

Overall, the main contribution of this note is to clarify that the CMH test can be understood not only through the traditional common-odds-ratio framework, but also as a test of the average treatment effect based on an estimator that is unbiased by randomization alone. This perspective connects a classical stratified analysis tool with modern causal inference. It also clarifies a practically important point: in stratified randomized trials with binary endpoints, the asymptotic elegance of unrestricted AIPW inference does not eliminate finite-sample concerns arising from sparsity, boundary estimates, and treatment-effect heterogeneity across strata. In such settings, the conditional nature and pooled variance structure of the CMH test can remain attractive for testing, even when AIPW methods are preferred for point estimation or for other estimands. 

\section*{Acknowledgements}

I thank my colleagues Dominic Magirr and Lillian Yau for helpful discussions about the topic. AI tools, including ChatGPT and Claude Opus, were used during the preparation of this manuscript to identify relevant literature, to suggest and check derivations and arguments, to write R code for implementation and simulation, to draft preliminary text, and for stylistic revision. All suggestions were checked by the author. The conceptual framing, final selection of arguments, interpretation of results, and responsibility for the final manuscript remain with the author.

\bibliographystyle{plainnat}
\bibliography{references}

\end{document}